\documentclass[lettersize,journal]{IEEEtran}

\usepackage{amsmath,amsfonts}
\usepackage{algorithmic}
\usepackage{algorithm}
\usepackage{array}
\usepackage[caption=false,font=normalsize,labelfont=sf,textfont=sf]{subfig}
\usepackage{textcomp}
\usepackage{stfloats}
\usepackage{url}
\usepackage{verbatim}
\usepackage{graphicx}
\usepackage{cite}
\usepackage{siunitx} 
\usepackage{xcolor}
\usepackage{authblk}

\newcommand{\um}{\si{\micro\meter}}
\newcommand{\us}{\si{\micro\second}}
\newcommand{\uBq}{\si{\micro\becquerel}}
\newcommand{\uBqpkg}{\si{\micro\becquerel/\kilogram}}
\newcommand{\mBqpkg}{\si{\milli\becquerel/\kilogram}}
\newcommand{\Am}[1]{$^{#1}$Am}
\newcommand{\Pu}[1]{$^{#1}$Pu}
\newcommand{\Ra}[1]{$^{#1}$Ra}
\newcommand{\Th}[1]{$^{#1}$Th}
\newcommand{\Tl}[1]{$^{#1}$Tl}
\newcommand{\U}[1]{$^{#1}$U}
\newcommand{\Rn}[1]{$^{#1}$Rn}
\newcommand{\Bi}[1]{$^{#1}$Bi}
\newcommand{\Po}[1]{$^{#1}$Po}
\newcommand{\Pb}[1]{$^{#1}$Pb}
\newcommand{\Gd}[1]{$^{#1}$Gd}
\newcommand{\Cm}[1]{$^{#1}$Cm}
\newcommand{\zerodbd}{$0\nu\beta\beta$}
\begin{document}

\title{Development of decay energy spectroscopy for radio impurity analysis}

\author[1,2]{J.S. Chung \thanks{ Corresponding authors: Y.H. Kim (email: yhk@ibs.re.kr) and H.L. Kim (email: khl7984@gmail.com)}}
\author[1]{O. Gileva}
\author[2]{C. Ha}
\author[1]{J.A Jeon}
\author[1]{H.B. Kim}
\author[1]{H.L. Kim }
\author[1,3]{Y.H. Kim  }
\author[1]{H.J. Kim}
\author[1]{M.B Kim}
\author[1,3]{D.H. Kwon}
\author[1]{ D.S. Leonard}
\author[1]{D.Y. Lee}
\author[1]{Y.C. Lee}
\author[1]{H.S. Lim}
\author[1,3]{K.R. Woo}
\author[1]{J.Y. Yang}

\affil[1]{Center for Underground Physics, Institute for Basic Science (IBS), Daejeon 34047, Republic of Korea}
\affil[2]{Department of Physics, Chung-Ang University, Seoul 06973, Republic of Korea} 
\affil[3]{IBS School, University of Science and Technology (UST), Daejeon 34113, Republic of Korea}

\markboth{Journal of \LaTeX\ Class Files,~Vol.~14, No.~8, August~2021}%
{Shell \MakeLowercase{\textit{et al.}}: A Sample Article Using IEEEtran.cls for IEEE Journals}

\maketitle

\begin{abstract}
We present the development of a decay energy spectroscopy (DES) method for the analysis of radioactive impurities using magnetic microcalorimeters (MMCs). The DES system was designed to analyze radionuclides, such as \Ra{226}, \Th{228}, and their daughter nuclides, in materials like copper, commonly used in rare-event search experiments. We tested the DES system with a gold foil absorber measuring 20$\times$20$\times$0.05 mm$^3$, large enough to accommodate a significant drop of source solution. Using this large absorber and an MMC sensor, we conducted a long-term measurement over ten days of live time, requiring 11 ADR cooling cycles. The combined spectrum achieved an energy resolution of 45\,keV FWHM, sufficient to identify most alpha and DES peaks of interest. Specific decay events from radionuclide contaminants in the absorber were identified. This experiment confirms the  capability of the DES system to measure alpha decay chains of \Ra{226} and \Th{228}, offering a promising method for radio-impurity evaluation in ultra-low background experiments. 
\end{abstract}


\section{Introduction}
\IEEEPARstart{D}{ecay} energy spectroscopy (DES) has emerged as a precise radiometric technique for radionuclide analysis~\cite{koehler2021as}. By encapsulating source materials in a gold absorber, surrounding the source with a solid angle of 4$\pi$, DES measures the total energy deposited from each radioactive decay. High-resolution microcalorimeters, such as magnetic microcalorimeters (MMCs) or transition-edge sensors (TESs), are utilized to achieve a precise measurement of the energy of these decay events in the absorber.

The principle of DES for alpha emitters was first suggested and demonstrated in Ref.~\cite{sjlee2010jpg}, which highlighted the distinct difference between the decay energy of alpha particles from \Am{241} enclosed in a gold 4$\pi$ absorber and the energy of alpha particles from an external \Am{241} source. This method was subsequently adapted for DES measurements of a \Ra{226} sample using a gold absorber and a microfabricated MMC as the sensor technology~\cite{ysjang2012ari}. The resulting decay-energy peaks corresponded to \Ra{226} and its daughter nuclides. 
TESs were also utilized in DES, leading to significant improvements in energy resolution and spectral shape~\cite{croce2016jltp}. A high-resolution DES spectrum was obtained from a plutonium reference material, with clear identifications of \Pu{239} and \Pu{240} at their respective decay-energy values. The DES peaks of alpha decay were nearly Gaussian in shape, with a resolution of 1.3\,keV FWHM. Recently, similar DES performance has been obtained with an MMC for nuclear material analysis~\cite{gbkim2024arxiv}.

On the other hand, the DES method has been effectively employed for the accurate measurement of various beta spectra~\cite{loidl2010ari,loidl2014ari,paulsen2019jinst}. In conventional solid-state detectors, significant energy loss occurs within the source sample and in inactive regions of the detector, which distorts the true spectral shape of the electron energy emitted from beta decays. Alternative gas and liquid detectors face challenges with energy resolution and detection efficiency. However, in the source-equal-to-detector configuration of DES, using 4$\pi$ gold absorbers, the spectral shape of beta decays can be measured accurately without energy loss while allowing for high resolution detection with microcalorimeters.

Since detection efficiency for DES is virtually 100\% at the count rates of interest~\cite{kavner2024nima,schulze2024mst,kossert2023ari}, it can be utilized for absolute decay counting and primary radioactivity standardization. Precise activity counting can be achieved by using an absorber that encloses a known quantity of the source sample (e.g., a drop-dried sample from a solution). However, accurate activity counting also requires precise control and careful measurement of the source solution quantity~\cite{schulze2024mst}.

We aim to develop a DES method to evaluate radioimpurities in the clean materials used in rare-event search experiments. For example, all cryogenic particle detectors used for direct dark matter detection and neutrinoless double beta decay (\zerodbd) searches utilize copper to cool the detectors to mK temperatures and shield them from environmental gamma backgrounds. It is critical to minimize the influence on the detector from radiation background from the copper itself. However, there are technical challenges in investigating the activity levels of isotopes such as \Ra{226} and \Th{228} in copper.

While thorium and uranium impurities (mainly \U{238} and \Th{232}) can be measured using ICP-MS analysis after chemical extraction~\cite{grinberg_determination_2005, laferriere_novel_2015,arnquist_automated_2020,gileva2023ari}, their radioactive daughters are often not in secular equilibrium in pure materials~\cite{agrawal2024fp}. The isotopes \Bi{214} and \Tl{208}, which are particularly problematic for \zerodbd{} experiments, are associated with the decays of \Ra{226} and \Th{228}, but not directly with \U{238} and \Th{232}, respectively. In addition, due to the expected low activity levels in copper, the self-shielding of the sample material itself makes HPGe counting impractical for detecting \Ra{226} and \Th{228} impurities.

In this report, we characterized an MMC setup using a large gold absorber sizing 20$\times$20$\times$0.05\,mm$^3$. The absorber, which is two orders of magnitude larger than those typically used in DES with an expected heat capacity of 100\,nJ/K at 70\,mK, was selected to enclose a drop-dried sample containing radium and thorium compounds extracted from several kilograms of copper.

\begin{figure}[!t] %fig1
\centering
\includegraphics[width=0.95\columnwidth]{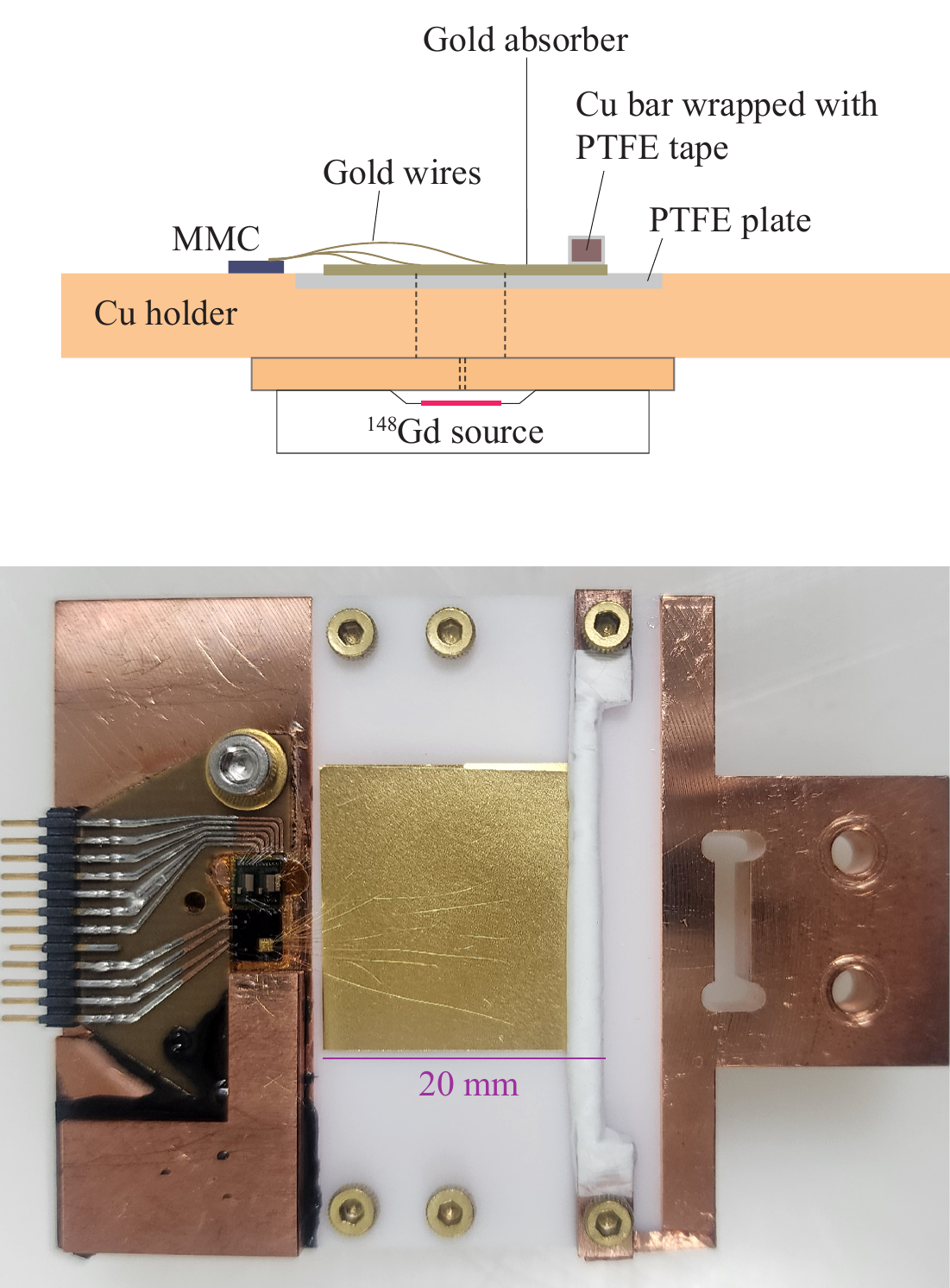}
\caption{
Schematic diagram of the experimental setup with a large gold absorber (\textit{top}). A \Gd{148} alpha source was placed at the bottom of the sample holder.  Photograph of the DES setup with the gold absorber (20$\times$20$\times$0.05\,mm$^3$) (\textit{bottom}). Gold wires were bonded between the MMC sensor and the gold absorber for thermal connection. A copper bar wrapped with several layers of PTFE tape was used to hold the foil in place on a PTFE plate. }
\label{fig:setup}
\end{figure}

\section{Experiment}
To assess the performance of the DES setup and verify the level of radio impurities in the absorber material, we tested a large gold foil as an absorber without depositing any source material.
The gold absorber we used was first cut into a 40$\times$20~mm$^2$ section from commercially available 25~\um{} thick foil. After undergoing a cleaning process—dipping the foil in 5\% nitric acid for 5~minutes, followed by rinsing with highly deionized water, it was folded in half and diffusion welded at \SI{600}{\celsius} for 3~hours in a vacuum oven. Two quartz plates were used to press the foil with a 800\,g copper weight.

A DES setup was constructed with an MMC sensor and the absorber foil, formed in a  20$\times$20$\times$0.05~mm$^3$ structure, which served as a dummy sample without intentional source material inside (as shown in Fig. \ref{fig:setup}). The thermal connection between the MMC sensor layer and the gold absorber was established using ten 25\,\um{} diameter gold bonding wires distributed across the foil~\cite{ysjang2012ari,wsyoon2012jltp}.

We utilized one of the MMCs developed for the early phase of the AMoRE \zerodbd{} experiment~\cite{cskang2017sust,amore2019epjc}. It was fabricated from a Au:Er alloy batch with 1000~ppm Er for the pilot stage of the experiment. The sensor layer was 3~\um{} thick and  1$\times$1~mm$^2$ wide, and fabricated on a double-sided meander-shaped coil with each side having 32 nH inductance.  
The heat capacity and magnetization properties of the MMC sensors in this batch have been well-studied at low temperatures~\cite{sgkim2021ieeetas}. The meander coil of the MMC setup was connected to a current-sensing SQUID having a 5.5\,nH self inductance and an input coupling of $\SI{3.5}{\micro\ampere /\Phi_0}$ with a pair of aluminium wires. The SQUID was designed and fabricated to match with the AMoRE MMCs  at The Kirchhoff Institute for Physics (KIP), Heidelberg University. 

Before conducting measurements with the large gold absorber, we first performed a test experiment using a small gold absorber and a similar detector setup, which included an MMC sensor and a 2$\times$2$\times$0.05\,\um{} gold foil. 

Both the small and large absorber experiments were carried out in an adiabatic demagnetization refrigerator (ADR) surrounded by 5 cm thick lead shielding. Given the goal of establishing low-activity measurements with this method, long-term data collection was necessary. Several cycles of magnetization and demagnetization were repeated for each experiment.

For energy calibration, a \Gd{148} source emitting 3183~keV alphas was used. To adjust the source rate of about 0.3\,cps, a copper collimator was installed, positioned on the opposite side to avoid interference with the gold wires connected to the MMC, as shown in the schematics of the experimental setup  in Fig.~\ref{fig:setup}. The 3183~keV energy from the \Gd{148} source is not within our region of interest (ROI), 4--8~MeV, for the radio impurity analysis.

\section{Results and Discussions}

\begin{figure}[!t] %fig2
\centering
\includegraphics[width=\columnwidth]{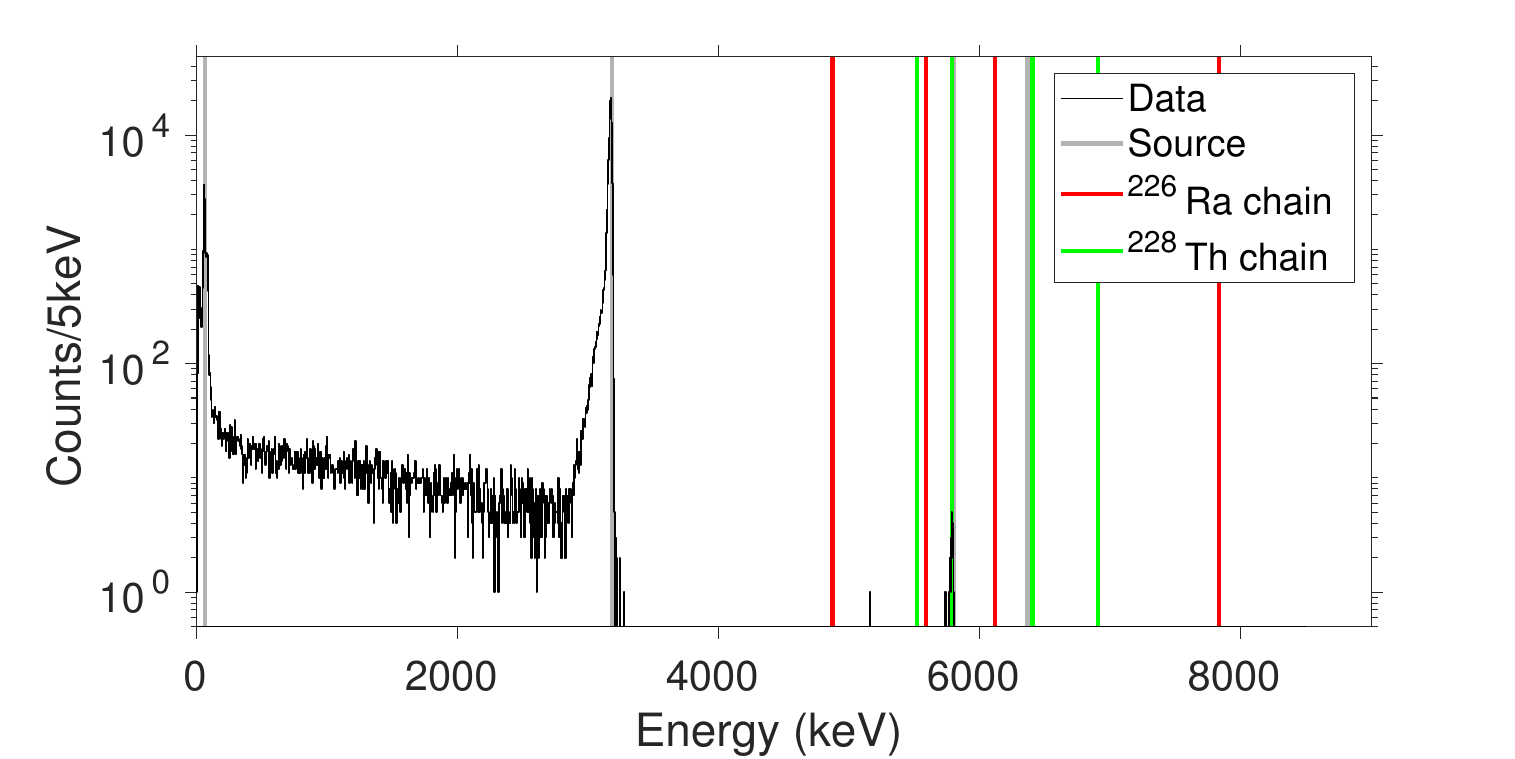}
\includegraphics[width=\columnwidth]{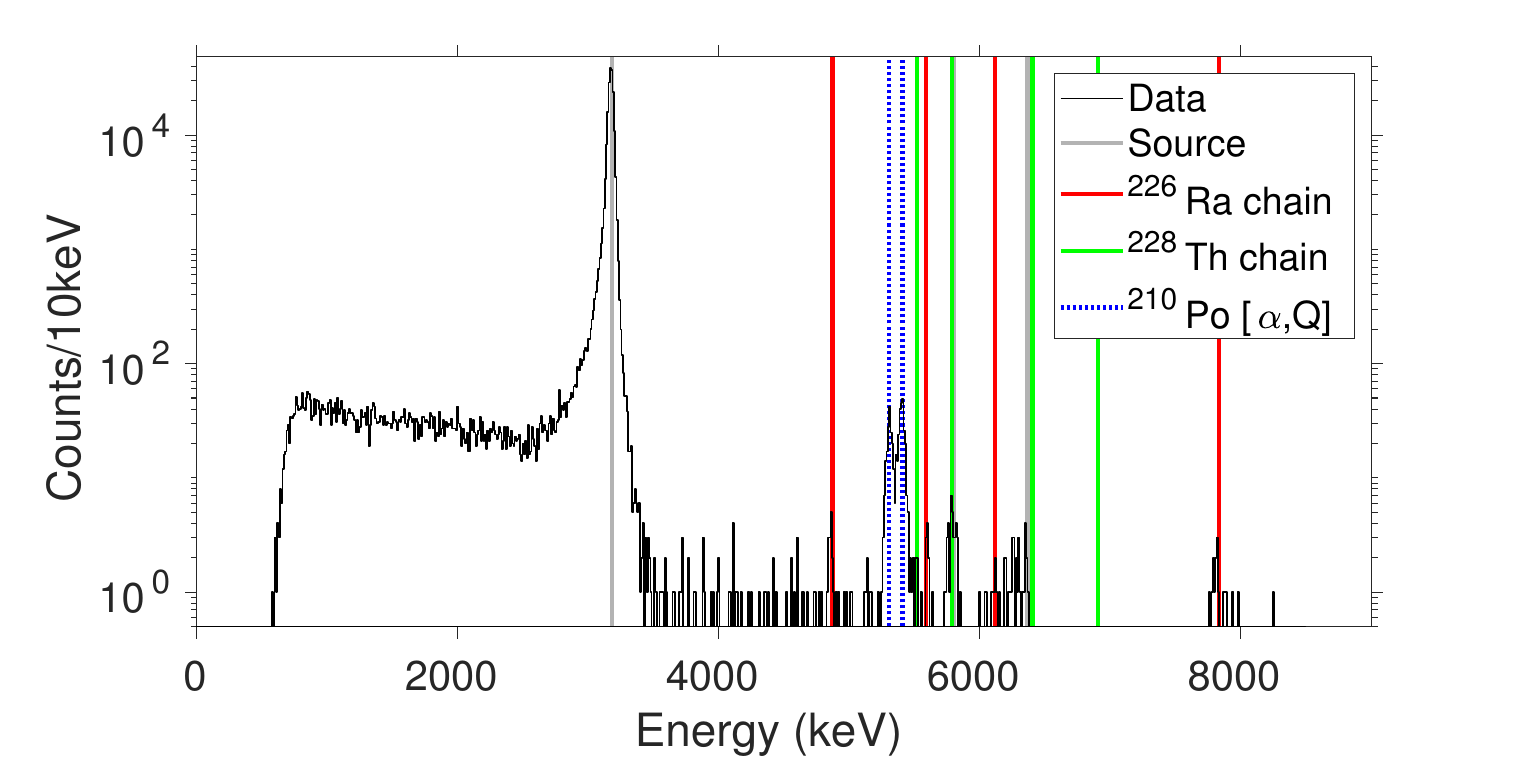}
\includegraphics[width=\columnwidth]{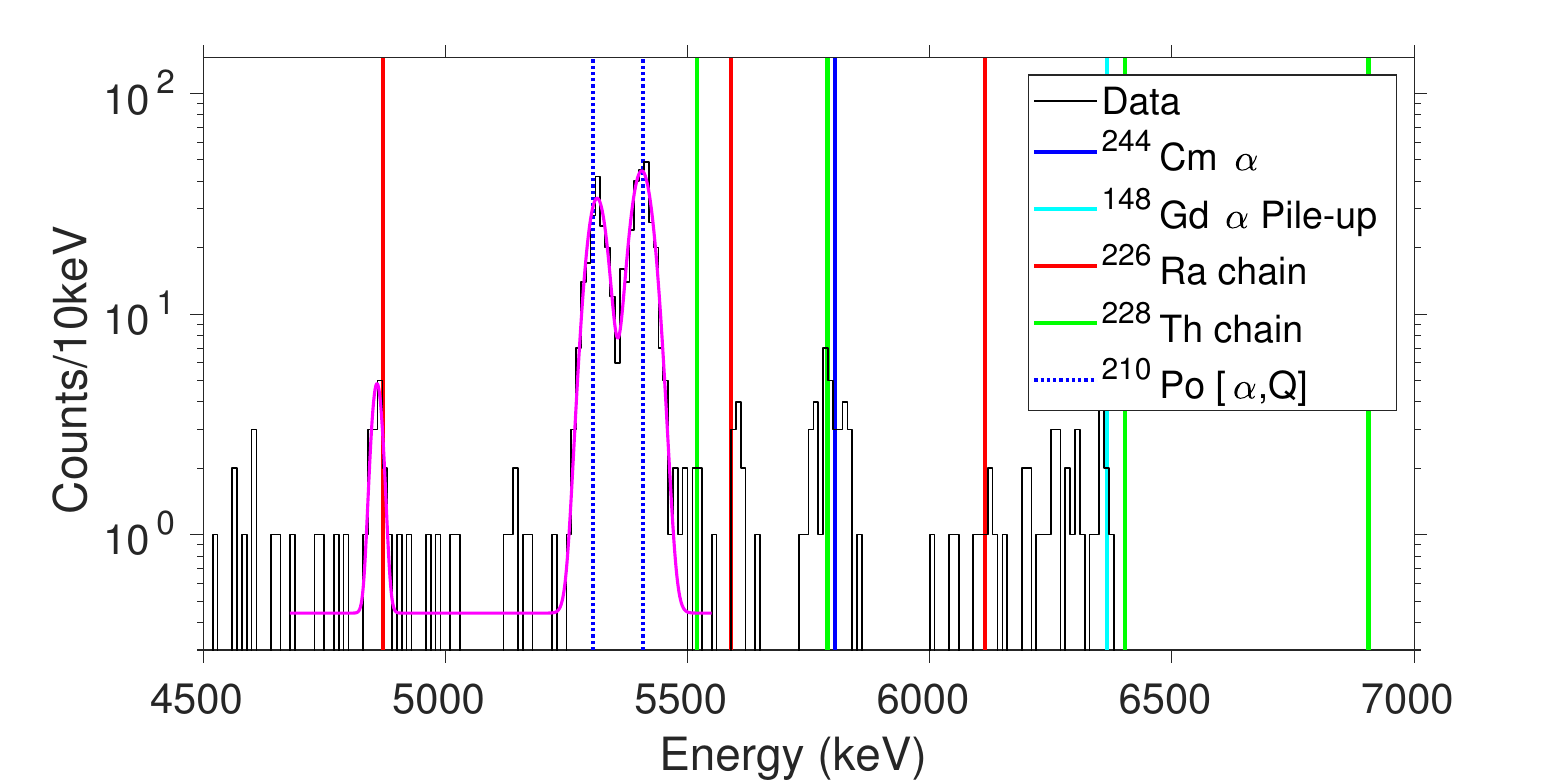}
\caption{Full energy spectrum measured with the small gold absorber for 2.3~day live time (\textit{top}). Full energy spectrum measured with the large gold absorber for 10.2 day live time (\textit{middle}). A zoomed spectrum in the energy region of interest with the large gold absorber (\textit{bottom}).}
\label{fig:spectra}
\end{figure}

Fig.~\ref{fig:spectra} shows the full energy spectra of the measured pulses for both the small and large absorbers, presented in the top and middle plots, respectively. 

For the measurement with the small absorber  
almost no events were detected in the spectrum other than the alpha peak at 3183\,keV.   
A one-point linear calibration was applied with this alpha peak.
However, above 5\,MeV, a group of additional events was observed. Upon investigation, we identified these events as originating from unexpected contamination by \Cm{244}, which has two major lines at 5763\,keV and 5805\,keV. Thanks to the detector's high energy linearity, the energy of these events aligned well with the expected \Cm{244} lines.

The vertical lines in all plots of Fig.~\ref{fig:spectra} mark possible peaks from the source and the decay chains of \Ra{226} and \Th{228}, grouped by different colors. The gray lines indicate peaks from the source, including unresolved pileup events from two accidental measurements of \Gd{148} alpha particles within the rise time of the signals. 

The spectrum in the top plot represents a combined result of 65\,mK measurements from three ADR cooling cycles with a total live time of 56 hours. The combined spectrum shows a full width at half maximum (FWHM) energy resolution of 20\,keV from a simple Gaussian fit, with some low-energy tail caused by alpha straggling in the source.

The result with the small absorber demonstrates that this source can be used as a calibration source in DES measurements of rare activities from the \Ra{226} and \Th{228} decay chains. It does not emit significant alpha radiation above 4\,MeV. The small activity of \Cm{244} may also aid in testing energy calibration and gain correction for long-term measurements.

The middle plot of Fig.~\ref{fig:spectra} shows the combined spectrum of eleven data sets taken using the large absorber, with a total live time of 244 hours at 65\,mK. The primary peak from the \Gd{148} alpha events exhibits a resolution of 45\,keV FWHM, determined by a simple Gaussian fit. 
We applied an optimal filtering method to determine the pulse amplitude~\cite{mccammon2005,yuryev2011,didomizio2011jinst}. Applying the same linear calibration using the 3183\,keV alpha peak, an excellent energy linearity was demonstrated with energy differences of less than 10\,keV for the \Cm{244} and \Po{210} alpha peaks. The relative rates of the two alpha peaks, \Gd{148} and \Cm{244}, are consistent at $2\times10^{-4}$ for both cases, within the statistical error of the total counts for \Cm{244}. A greater number of events are observed above 4\,MeV with the large gold absorber compared to the smaller one. Four DES peaks were identified from the \Ra{226} decay chain:
three from the decay sequence \Ra{226} $\rightarrow$ \Rn{222} $\rightarrow$ \Po{218} $\rightarrow$ \Pb{214} and another from \Bi{214}-\Po{214} $\rightarrow$ \Pb{210} decay. The consecutive beta-alpha decays of \Bi{214}-\Po{214} are considered as a single event because the half life of \Bi{214}, 164\,\us, is much shorter than the rise time of the signals, which was about 1~ms.  

Two peaks were identified related to \Po{210}: one for the alpha peak and another for the DES peak from \Po{210} decay. The DES peak likely resulted from \Po{210} alpha decays occurring inside the absorber, while the \Po{210} alpha peak may have been caused by \Po{210} contamination near the gold foil absorber. The spectrum also shows a number of background events scattered across a wide energy range, ungrouped to any specific energy. These events are also likely alpha backgrounds from contamination of alpha emitters near the gold absorber.

The bottom plot is a zoomed-in view of the ROI from the middle plot. We performed an unbinned maximum likelihood fit to evaluate the decay rates of \Ra{226} and \Po{210} impurities in the sample absorber itself.  The magenta curve represents the fit result, which includes three Gaussian functions and a flat background with a total of 10 free parameters. The activity of \Ra{226} in the gold foil is determined to be 14$\pm$4\,\uBq{}, accounting for the 3.6\% emission probability of 186\,keV gamma rays~\cite{ysjang2012ari}, while the \Po{210} activity is 191$\pm$16\,\uBq. The DES peaks of \Rn{222} and \Po{218}, daughter isotopes of \Ra{226}, are also observed in this spectrum.
 However, their rates may not be in equilibrium with \Ra{226} because the measurement period was not long enough after the sample preparation, compared to the 3.8-day half-life of \Rn{222}, an inert gas that may have escaped from contaminants on the gold surface before they were folded into the foil.
Some unresolved pileup events are seen at and below the gray line at 6365 keV, which corresponds to twice the \Gd{148} alpha energy. These pileups may interfere with \Po{218} events, introducing uncertainty in determining the \Po{218} activity.
Furthermore, no events were detected near the 6906\,keV line, the decay energy of \Po{216}, indicating that no decay sequences associated with \Th{228} decay occurred in the gold absorber during the measurement, with a sensitivity limit of 1.1\,\uBq.

Despite the large heat capacity of the gold absorber, the spectrum has reasonable energy resolution. 
The resolution obtained with the gold absorber is sufficient for practical DES analysis. The \Ra{226} and \Th{228} decay series feature distinct DES lines, such as the 4871\,keV line of \Ra{226} and the 6906\,keV line of \Po{216}, which do not overlap with others. If background rates are not dominant, a resolution of 50 keV FWHM is sufficient to identify these specific lines, allowing the determination of \Ra{226} and \Th{228} activity rates in the sample. Additionally, background reduction can be enhanced by applying time-coincidence analysis, as the half-lives of \Po{218} and \Rn{220} are 3.1 minutes and 55 seconds, respectively.

The signals with the small and large absorbers had rise times (i.e., the time elapsed between 10\%–90\% of the pulse height on the rising edge) of 0.2 ms and 1.2 ms at 65 mK, respectively. These rise time values exhibited slight temperature dependence in 50--70 mK. From previous studies~\cite{wsyoon2012jltp,ysjang2012ari,iwkim2017sust}, the characteristic time constant, $\tau_\mathrm{rise}$, between the absorber and the sensor is understood as 
 $\tau_\mathrm{rise}=C_\mathrm{eff}/G_\mathrm{as}$ where $C_\mathrm{eff}$ is the effective heat capacity of the two components and $G_\mathrm{as}$ is the thermal conductance. 
 The measured rise time for the smaller absorber agreed well with the calculated value of $\tau_\mathrm{rise}$, but this simple estimation did not hold for the signals from the larger absorber.

For a flat foil with a large area, thermal diffusion in the lateral direction must be considered when determining the time constant for thermal equilibrium. This time constant, $\tau_\mathrm{foil}$, for a square foil with side length $l$, can be expressed as $\tau_\mathrm{foil} = \gamma \rho l^2 /L_0$, where $\gamma$ is the Sommerfeld coefficient of the electronic specific heat, $\rho$ is the resistivity, and $L_0$ is the Lorentz number. 
Note that $\tau_\mathrm{foil}$ is temperature-independent because the specific heat and thermal conductivity of gold are proportional to temperature at low temperatures.
For a gold foil with the given dimensions of the large absorber and a residual resistance ratio (RRR) of 20, this value is approximately 1\,ms. This diffusion time should be related to the measured time constant. 

While a more realistic model simulation of heat transfer for the given geometry and thermal properties would be necessary to fully reproduce the measured pulse shape, we infer that the size of the gold foil played a significant role in the rise time for the signals with the large absorber. To reduce the rates from environmental sources and the unresolved pileups, adjusting the foil size may be required.

\section{Conclusion}

In this study, we successfully developed a DES system with high sensitivity for radionuclide analysis, utilizing a large gold absorber coupled with an MMC sensor. The DES system is suitable for low-activity measurements associated with the decay chains of \Ra{226} and \Th{228} in source samples.
A 10-day DES measurement revealed an internal \Ra{226} activity of 14$\pm$4\,\uBq{} and no detectable \Th{228} activity, with a sensitivity limit of 1.1\,\uBq{} for a gold absorber with dimensions of 20$\times$20$\times$0.05\,mm$^3$.

Ref.~\cite{agrawal2024fp,gileva2023ari} reports recent ICP-MS measumrements of uranium and thorium impurities in high-purity copper samples  used in the AMoRE \zerodbd{} experiment, with values of 0.33$\pm$0.12\,pg of \U{238} and 0.26$\pm$0.11\,pg of \Th{232} per gram of copper for the ``2021'' NOSV Aurubis product. Assuming decay equilibrium (though not the case in practice), the equivalent activities are 4.1$\pm$1.5\,\uBqpkg{} for \Ra{226} and 1.1$\pm$0.5\,\uBqpkg{} for \Th{228}.
On the other hand, routine HPGe counting measurements provided detection limits of 0.49\,\mBqpkg{} and 1\,\mBqpkg{} for \Ra{226} and \Th{228}, respectively, with a more extensive effort in a similar sample still only reaching limits on the order of 100 \uBqpkg.

Therefore, applying this DES method to a gold foil with the drop-dried method for a solution containing extracted radium and/or thorium compounds from a kilogram of copper could yield far better sensitivities than those achieved by HPGe. Furthermore, a blank foil can be measured for its internal activities before source deposition, allowing efficient detection of radio impurities by comparing the activities before and after deposition.

In addition, with a longer data-taking time, the DES sensitivities can surpass the equivalent activity values derived from ICP-MS measurements. For instance, a DES setup installed in a dilution refrigerator (DR), which operates continuously at lower temperatures than the present ADR setup, could further enhance sensitivity.

In summary, decay energy spectroscopy shows great promise for analyzing \Ra{226} and \Th{228} activity levels in high-purity copper and potentially other pure materials used in
ultra-low background experiments. It can serve as a complementary tool alongside, or even surpass the detection limits of, other radiochemical analysis techniques such as ICP-MS and HPGe.

\section*{Acknowledgments}
This research is supported by Grant no. IBS-R016-A2.
JSC acknowledges the support by the Chung-Ang University Graduate Research Scholarship in 2023.

\vfill

\end{document}